\documentclass[sigconf, screen, authorversion]{acmart}
\usepackage[utf8]{inputenc}
\usepackage{booktabs}
\usepackage{multirow}
\usepackage{microtype}
\usepackage{wasysym}

\copyrightyear{2022}
\acmYear{2022}
\setcopyright{acmlicensed}
\acmConference[AIES '22] {Proceedings of the 2022 AAAI/ACM Conference on AI, Ethics, and Society}{August 1--3, 2022}{Oxford, United Kingdom.}
\acmBooktitle{Proceedings of the 2022 AAAI/ACM Conference on AI, Ethics, and Society (AIES '22), August 1--3, 2022, Oxford, United Kingdom}
\acmPrice{15.00}
\acmISBN{978-1-4503-9247-1/22/08}
\acmDOI{10.1145/3514094.3534186}

\begin{document}
\fancyhead{}

\author{William Seymour}
\email{william.1.seymour@kcl.ac.uk}
\orcid{0002-0256-6740}
\affiliation{%
  \institution{King's College London}
  \city{London}
  \state{UK}
}
\affiliation{%
  \institution{University of Oxford}
  \city{Oxford}
  \state{UK}
}

\author{Max Van Kleek}
\email{max.van.kleek@cs.ox.ac.uk}
\affiliation{%
  \institution{University of Oxford}
  \city{Oxford}
  \state{UK}
}

\author{Reuben Binns}
\email{reuben.binns@cs.ox.ac.uk}
\affiliation{%
  \institution{University of Oxford}
  \city{Oxford}
  \state{UK}
}

\author{Dave Murray-Rust}
\email{d.s.murray-rust@tudelft.nl}
\orcid{0000-0001-6098-7861}
\affiliation{%
  \institution{TU Delft}
  \city{Delft}
  \state{The Netherlands}
}

\title[Respect as a Lens for the Design of AI Systems]{Respect as a Lens for the Design of AI Systems}


\begin{abstract}
Critical examinations of AI systems often apply principles such as fairness, justice, accountability, and safety, which is reflected in AI regulations such as the EU AI Act. Are such principles sufficient to promote the design of systems that support human flourishing? Even if a system is in some sense fair, just, or `safe', it can nonetheless be exploitative, coercive, inconvenient, or otherwise conflict with cultural, individual, or social values. This paper proposes a dimension of interactional ethics thus far overlooked: the ways AI systems should treat human beings. For this purpose, we explore the philosophical concept of respect: if respect is something everyone needs and deserves, shouldn't technology aim to be respectful? Despite its intuitive simplicity, respect in philosophy is a complex concept with many disparate senses. Like fairness or justice, respect can characterise how people deserve to be treated; but rather than relating primarily to the distribution of benefits or punishments, respect relates to how people regard one another, and how this translates to perception, treatment, and behaviour. We explore respect broadly across several literatures, synthesising perspectives on respect from Kantian, post-Kantian, dramaturgical, and agential realist design perspectives with a goal of drawing together a view of what respect could mean for AI. In so doing, we identify ways that respect may guide us towards more sociable artefacts that ethically and inclusively honour and recognise humans using the rich social language that we have evolved to interact with one another every day. 
\end{abstract}

\maketitle

\section{Introduction}
AI systems are being adopted in a growing number of contexts, fed by an ever-increasing amount of data. These systems shape people's lives by filtering and ranking the information they see, and classifying them into groups based on different feature spaces. These technologies are simultaneously becoming more adaptive and human-like, capable of responding to people using natural affordances such as spoken dialog and gesture. This means that people often expect them to act as humans would, ascribing to them human-like social capabilities~\cite{reeves1996media}. Their use in home and communications contexts also means that AI systems increasingly come to mediate interactions \textit{between} people, changing the ways that we relate to each other, thus becoming intertwined with the performance and experience of social practices.

Concerns have been raised over the ways that AI systems represent people (such as through performance of gender~\cite{west2019blush}), introduce and entrench problematic biases (such as with recidivism prediction~\cite{angwin2016machine}), and negatively shape interpersonal relationships (such as with social media~\cite{10.1145/3313831.3376482}). In other areas of research there have been efforts to take existing moral concepts and operationalise them for use in AI systems, a notable example being the emerging body of work on fairness in machine learning\footnote{\url{https://facctconference.org}}). But explicitly codifying abstract ethical principles often leads, in turn, to abstract frameworks that are cold and impersonal; the desire to `fix' unethical engineering artefacts takes precedent over the need to treat individual \textit{people} properly.

To this end, this paper presents the concept of respect as a parallel design goal for AI systems, encompassing a broad range of aspects that concern how people are treated. Respect forms a person's measure of their own self-worth, and is a reflection of the value they place in others. Being treated with respect bolsters one's self-esteem and confidence, promoting positive mental health and happiness, while mutual respect builds robust social bonds and cultivates mutual trust. A lack of respect, meanwhile, can undermine a person's well-being in as many ways as being respected can benefit them, even disenfranchising them of their very personhood. The idea that every person is entitled to a basic level of respect as a human being is core to many cultures and is enshrined in international law in the form of human rights~\cite{dillon1992respect}.

We begin by giving a brief overview of respect as it relevant to HCI and AI systems, including how the ethical principles that underpin different perspectives apply to modern machine learning technologies. From this we show how respect can help us to navigate the complex ecologies of people and devices that AI systems are deployed in and how respectful AI practices can complement existing design methodologies. Finally, we explore how these practices apply across the entire system life cycle from design to decommissioning and build on work around individual and intersectional fairness to develop a notion of individual \textit{respect} that helps to re-frame issues around fairness and classification around individual people and their unique and varied histories.

\section{Background: A brief history of Respect}\label{sec:background}
We begin by exploring a series of responses to the question of what it means to be respectful. Drawing on examples from philosophy and the human sciences, we explore the relevance of thinking on respect to the design of contemporary AI systems in HCI. Above and beyond being a value to design for respect offers a way of structuring difficult discussions around ethics and AI systems, unpacking why different options may be (un)desirable. Respect is well placed to do this as a `thick' concept, describing attitudes and relationships, as well as containing within it explicitly normative dimensions~\cite{williams2011ethics}. Here, we start with the history given in \cite{sep-respect}, through Darwall, Hudson, and Kant, but then extend this to take in contemporary discourse around care respect and a broadening notion of where respect might apply---summarised in Table~\ref{tab:typesofrespect}.

\begin{table*}\centering
\caption{Summary of key perspectives on respect from various conceptualisations and taxonomies.}\label{tab:typesofrespect}
\small
\begin{tabular}{p{0.15\textwidth}p{0.13\textwidth}p{0.6\textwidth}}\toprule
\textbf{Source} &\textbf{Type} &\textbf{Description} \\\midrule
\multirow{2}{*}{\textbf{Darwall}~\cite{darwall1977two}} & Recognition & Changing one's behaviour in consideration of some fact about a situation. \\
& Appraisal & The appraisal of people against some standard of excellence, such as their accomplishments, characteristics, or perceived skill. \\
\midrule
\multirow{4}{*}{\textbf{Hudson}~\cite{hudson1980nature}} & Directive & Modifying behaviour to recognise the authority of a stated preference, law, rule, or principle. \\
& Obstacle & Adapting behaviour to manage something that is an obstacle to one's own goals. \\
& Institutional & Following the practices, customs, and hierarchies of society and its institutions, regardless of one's opinion of the people within them. \\
& Evaluative & Responding to how well people and objects meet certain standards. \\
\midrule
\textbf{Kant}~\cite{kant2002groundwork} & Categorical Imperative & Treating people genuinely, as \emph{ends} in of themselves rather than as a \emph{means} of accomplishing selfish objectives; recognising ``the status, worth, and individuality of others simply for being human, regardless of their place, position, or situation''. \\
\midrule
\textbf{Dillon}~\cite{dillon1992respect} & Care Respect & Recognising people for the individuals they are, responding to their needs, and promoting their well-being. \\
\midrule
\textbf{Mills}~\cite{mills2017blackradical} & Black Radical Kantianism & Acknowledging the ideological and psychological legacy of racism and the habits of disrespect that affect respect for others and self-respect. \\
\midrule
\multirow{2}{*}{\textbf{Goffman}~\cite{goffman1978presentation}} &Deference &Recognising another's place, role, status, and position; conveyance of appreciation or devotion, facework, or taking action to save others' reputation or embarrassment. \\
&Demeanour & Communication (through behaviour and presentation) of how one wishes to be seen and treated, and giving that treatment to others. \\
\midrule
\textbf{McDowell}~\cite{MCDOWELL2007276} & Respectability & External perceptions of socioeconomic status, supported by certain kinds of behaviour. \\
 &Deference &Actions that signal deference for others, whether based on compliance with authority or personal respect \\
 \midrule
\textbf{Schrimer}~\cite{schirmer2013respect} & Ascribed agency & Giving people the space they need to skillfully be themselves. \\
\bottomrule
\end{tabular}
\end{table*}

While respect is sometimes used to describe the appraisal of someone~\cite{darwall1977two}, such as their punctuality or skill as an artist, in the context of AI systems we are more concerned with respect as giving weight to a fact or characteristic when deciding how to act. Hudson developed a taxonomy of different types of respect that captures this, distinguishing between respect when following a law or rule (\textit{directive respect}), taking care of obstacles and dangers, such as adding handrails or backing away from a dangerous animal (\textit{obstacle respect}), or following the norms of an institution (\textit{institutional respect})~\cite{hudson1980nature}. 

While these definitions capture many aspects of respect as we use it in everyday life, in order to fully explore its relevance in the design of AI systems we now consider conceptualisations of respect that inform our social and cultural understanding of how we should treat each other as human beings. For this we begin by examining Kant's categorical imperative, before exploring subsequent criticism from work in feminist philosophy and the philosophy of race. This is followed by conceptualisations of respect as a property that emerges out of our interactions with people, objects, and constructs.

\subsection{Kant and the Categorical Imperative}
Respect became a foundational concept in Western moral philosophy during the Enlightenment when Immanuel Kant argued that there are certain essential moral duties, or obligations, that people owe to others unconditionally, regardless of context ~\cite{kant2002groundwork}. In contrast to the subjective moral frameworks of his day, Kant saw respect for others as part of one's \textit{``categorical imperative''} as a moral entity~\cite{kant2002groundwork}: that people should recognise the status, worth, and individuality of others simply for being human, regardless of their place, position, or situation, because all people are, themselves, moral agents. This duty imposed several requirements on how people treat others. Kant's \textit{humanity formula} describes the need to treat others genuinely; as ends in themselves rather than simply for one's own advantage. Doing this entails both negative and positive duties, such as avoiding and preventing the restriction of the freedom, autonomy or personhood of others~\cite{blau2015human}, and in other ways promoting the very same. This understanding of respect for persons forms the basis for modern human rights frameworks, operationalising the duty not to violate these rights which are due to ``all members of the human family''~\cite{assembly1948universal, dillon1992respect}. These frameworks encompass many economic, social, and cultural aspects seen as essential for ``a decent life''~\cite{blau2015human}. 

\subsection{Critiques of Kant: Feminist Philosophy and Philosophy of Race}
Kant's philosophy emerged from the latter stages of the Enlightenment in Europe, a period and place in which the personhood of vast swathes of humanity was not accepted by those `enlightened' thinkers. Unsurprisingly, Kant's ideas about what respect is and how it should be shown have subsequently been critiqued, developed, and transformed by a more diverse set of traditions of moral and political thought. While there are too many of these to mention, we present two strands of thought on respect which are in part responses to Kant, and are of potential relevance to the design of respectful AI systems.

Some accounts of respect emerge out of a critique of the way that Kantian respect is rooted in an individual's moral capacity, and is impersonal in the sense that it ``flenses the individual down to the bare bones of abstract personhood''~\cite{johnson1982ignoring}. In grounding the need to respect others on the basis of their abstract moral capacity, Kantian respect risks ignoring the personal, social, and cultural characteristics that make us unique, as well as the network of relationships between people~\cite{farley1993feminist}.

Feminist philosophers have therefore sought to ground the need for respect not in an individual's abstract rationality, but in terms of their concrete, human, individual selves, with interdependent wants and needs. Dillon's account of \textit{care respect} ``requires not so much refraining from interference''---as an anti-paternalistic Kantian respect might suggest---but rather ``recognizing our power to make and unmake each other as persons [...] caring for others by responding to their needs, promoting their well-being, and participating in the realization of their selves and their ends''~\cite{dillon1992respect}.

In addition to abstracting away a person's unique and particular nature, the Kantian notion of respect may also fail to attend to the ways in which respect is unequally distributed under conditions of structural oppression, such as patriarchy and white supremacy. Work in feminist philosophy and philosophy of race has critiqued Kantian liberal notions of respect for persons on these grounds. For instance, Manne's account of the logic of misogyny notes that while respect is an attitude of good will which we expect of each other, those expectations are asymmetric by gender: `women in relations of asymmetrical moral support with men have historically been required to show him moral respect'~\cite[p:xix]{manne2017down}. According to such perspectives, unequal demands of respect between genders helps maintain patriarchal dominance by controlling, policing and punishing women who challenge it.

Similarly, in the context of racial domination, Mills argues that Kantian respect as traditionally conceived is a form of `racial liberalism', which assumes a `generic colorless political subject'~\cite{mills2017black}. Others argue that it embeds racist assumptions about the scope of personhood~\cite{allais2016kant}. Such critiques of Kantian republican liberalism could be interpreted as a rejection of the Kantian paradigm entirely, or alternatively as calls to re-constitute it. Mills represents the latter, arguing that Kantian respect needs to be `de-racialised'~\cite{mills2017blackradical}, and re-shaped to explicitly address how self-respect and respect for others ``will have been affected by race (as racism), leaving an ideological and psychological legacy, habits of disrespect, that will shape the `inclinations' most likely to be determinative and most imperatively to be resisted''~\cite{mills2017blackradical}.

\subsection{Respect as an Emergent Social Property}
Another challenge to Kantian notions of respect is to look at it not as a static, `extra-societal' property, but as something that emerges from our social behaviours and interactions. Goffman's classic treatise on presentation of self~\cite{goffman1978presentation} fundamentally presumes that all people have a moral obligation to treat others in ``appropriate ways'' as dictated by their social characteristics (as opposed to their innate personhood). While not explicitly using the term \emph{respect}, he develops the view that the way one acts serves the critical roles of both communicating and reinforcing others' views that one possesses one's social characteristics. This \emph{dramaturlogical} view constitutes the world of social actions as being---or simulating---theatrical performance, and explains people's actions in terms of how they serve various purposes for a particular time, place, and audience.

Goffman's \textit{deference} describes our symbolic ways of showing respect to others, such as through presentational rituals and social conventions; while \textit{demeanour} covers how people indicate their social characteristics or 'face', thereby exerting a moral obligation upon others treat them appropriately. McDowell similarly talks of deference and respectability~\cite{MCDOWELL2007276}. While deference often follows asymmetric hierarchies of power, Goffman gives the example of hospital staff, where doctors, nurses, and specialists should all defer to each other's expertise in their particular area. This connects to respect not just with regard to the idea of respecting one's particular authority or role, but that the appropriate and necessary ways of respecting may be asymmetric. Demeanour is then part of how a person handles their place within this network, with \textit{face-work} required in order to keep one's actions consistent with the social expectations established both by one's own demeanour and the deference due to others~\cite{goffman1955facework}.

Another key way that respect arises from our interactions with each other is in the ascription of agency to others. Constructivist views ground respect in \emph{ascribed agency}: ``showing respect towards other persons is suggested to mean the symbolic, communicative act of giving their agency the elbowroom they need in order to control their environment in a skillful, knowledgeable and independent manner''~\cite{schirmer2013respect}. Such a view acknowledges that there may be many factors that shape a person's actions that are out of their immediate control, whether psychological, social, cultural, environmental or physical. On hearing an insulting remark from a person with a different cultural background, one may consider whether it is a cultural mishap, and hence not attributed to the speaker's agency, or whether to ascribe agency and take the insult as an intended slight. Ascribed agency can also be a means of differentiating the interpretation of respectful behaviour from considerate or polite behaviour. In particular, the extent to which a polite or considerate action confirms or constrains a person's agency is said to define whether it is also respectful~\cite{schirmer2013respect}; offering an older adult a seat on public transport may be considerate, but could also be taken as disrespectful as it implies a lack of capability and agency. Similarly, guiding a visually impaired person without asking can be seen as considerate in supporting their needs, but shows a lack of respect through diminishing their agency in the situation.

\subsection{Respect and Non-human Things}
The treatments of respect examined so far all start from the idea that while respect can be for objects or concepts, it flows \emph{from} a person. In contrast, a new materialist perspective implies that respect can be enacted between a variety of human and non-human things rather than solely by humans as moral agents. This is vital for analysing human-computer interactions, as we find ourselves dealing with a range of devices, systems, and networks that have varying degrees and appearances of agency, allowing for different senses of respect and respectful behaviour. While there is deep debate about whether AI systems are truly intelligent ~\cite{gunkel2018robot,coeckelbergh2010robot,bryson2017and,birhane2020robot}, considering e.g. AI--driven voice assistants, it is clear that they \emph{perform} enough social functioning to be seen as respectful or disrespectful. 

A useful touchpoint here is Winner's look at the work of Robert Moses, \textit{``Do artifacts have politics?''}~\cite{winner1980artifacts}, in particular the `racist bridges' that did not afford enough clearance for busses, hence restricting parts of the city to only those wealthy enough to own cars. On a purely physical level, the bridge is relatively inert, and hard to see as a political actor. However, it is also very clearly the result of a collection of political decisions designed to entrench certain biases and hierarchies, and in this sense, can be ascribed a political character. In the same way, we might ascribe moral characteristics like respect and disrespect to the behaviour of technological artefacts, particularly those we interact with. There is a need to bring together perspectives on how technology shapes and mediates individual actions with views on how these micro-scale events relate to larger networks and concepts.
Working around hostile architecture~\cite{petty2016london}, Rosenberger develops a viewpoint that brings together accounts of networks and of interactions~\cite{rosenberger2014MultistabilityAgency}.

\subsection{Respect Beyond the Interpersonal}
Continuing this strand, the notion that respect is something ascribed and situational can be developed through ways of looking at the world that include the agency not just of humans but a broader range of entities including AI systems. At the risk of glossing a vast swathe of of thought, two concepts from the broad area of new materialism~\cite{coole2010introducing,alaimo2008material} are particularly generative with regards to respect:
\begin{itemize}
    \item Humans are not the only things that have agency, and hence the ability to show or receive respect;
    \item The world can be divided and compartmentalised in various ways, and the choice of division determines the phenomena observed.
\end{itemize}
To move closer to the source, concepts---such as respect---exist through ``specific agential intra-actions that the boundaries and properties of the `components' of phenomena become determinate and that particular embodied concepts become meaningful''~\cite{barad2007meeting}. In line with the previous section, respect can be seen an expressive performance: ``To convey respect entails finding the words, the gestures, and the layout of the physical space that makes respect felt and persuasive'' ~\cite{keevers2012social}. Similar concepts such as dignity are seen as a product of both a person's actions and their treatment by other people and social structures~\cite{glenn2010dignity}. Thus far the conceptualisations of respect presented have all been necessarily interpersonal, or concerned with people respecting people, but the new materialist perspective allows us to more fully consider respect for and by people, objects, and social constructs. This may initially appear far-fetched, but respect is commonly applied to many non-human things: showing directive respect for the laws of the land, or institutional respect for the `office of the president'; a recent anthropological study found respect-like concepts\footnote{Specifically, \emph{wayyuu}, a tacit moral model of respect and sacredness that is central to institutions and norms among the Arsi Oromo of Ethiopia~\cite{ostebo2018CanRespect}} applied to familial relations, significant and practical objects, areas of the home and particular foodstuffs~\cite{ostebo2018CanRespect}. 

\subsection*{Summary}
In this section we have summarised a few key ideas from a large and interdisciplinary debate on respect that capture the essence of respect as it relates to the design of AI systems. Starting by focusing on Kant's categorical imperative, that frames respect as something that exists separately from human construction, and its subsequent critiques, the section also included arguments that respect can emerge out of our interactions with other people, objects, and constructs. We now take these concepts and use them to demonstrate how respect can frame ethical elements of systems design, before developing them further into a framework around the different ways that respect can be built into AI systems.

\section{Respect and AI Design}
\subsection{What Would A Kantian AI Look Like?}
The initial Kantian understanding of respect given above is already useful for a number of reasons. It succinctly encapsulates an intuitive notion of respect, provides a means to operationalise it (``act as if the maxims of your action were to become through your will a universal law of nature''~\cite{kant2002groundwork}), and offers a way of evaluating systems that is often hard to pin down when using concepts like `fairness' or `privacy', particularly in their mathematically formalised forms. A system might be statistically `fair', for example, in that it treats all groups in a way that, while equal, denies them all personhood or violates their human rights (such as portrayed in Keyes et al.'s satirical speculative fiction, \emph{A Mulching Proposal}~\cite{keyes2019mulching}). Shaping decisions to the end of having a fair model might itself fail to treat individual's decisions as ends in themselves.

Kantian notions of respect for persons have historically been at the centre of the tradition of \emph{deontological} ethics, concerned with guiding and assessing the morality of an agent's choices---in terms of what is required, permissible, or forbidden in an absolute sense---rather than assessing the states of affairs those choices bring about as per \emph{consequentialist} ethics. Such consequentialist framings are implicit in approaches to fairness in machine learning which define fairness in terms of achieving an optimal balance of positive or negative outcomes and/or errors between protected groups.\footnote{Although, for a `deontological' approach to fairness see \cite{wang2020deontological}.} These approaches are often less grounded in considerations of whether it is acceptable for an AI system to treat a particular person in a certain way; instead, aiming for optimal group-level distributions. By instead focusing on the absolute moral permissibly of \emph{acts} regardless of consequences, deontological approaches eschew the calculation of outcomes as a method of ethical evaluation.

The categorical imperative places an obligation on designers to create systems that consider user's needs genuinely, rather than exploiting them as a means to an end. A primary example of this is the exploitation of user data under surveillance capitalism~\cite{zuboff2019age}, where systems with the primary purpose of harvesting users' personal data would be considered disrespectful and thus unethical under a Kantian view, because they treat users purely as a means of accruing value. Treating others as ends in themselves requires protecting their human rights, a position useful in establishing a normative framing for critiquing systems that, either directly or indirectly, disenfranchise people of such rights. This is particularly relevant in current discussions centring around whether AI-enabled and facilitated harms warrant regulation in the EU and elsewhere~\cite{smuha2021race}, because it grounds such violations in the unconditional obligations people have to respect others.

The notion of supporting the freedom of others as rational beings to think and act for themselves is a dominant aspect of Kant's ethical theory, and this has been interpreted by some as an obligation to support the autonomy of others, in terms of both thinking and action. Examples where this is violated include systems that try to coerce individuals to make certain choices, either through the use of ``dark patterns''~\cite{gray2018dark} that are already highly pervasive across the e-commerce ecosystem~\cite{mathur2019dark} or paternalistic notions that being guided by AI means platforms know what's best for users. This anti-paternalistic stance is not easily captured by other principles of user-centred design and ethical AI, such as user needs, fairness, privacy, transparency and accountability. Kantian accounts of respect require us to treat each other as rational agents; even if we think users are mistaken about their own best interests, that doesn't justify attempting to override their explicit, autonomous choice to reject an AI system's output.

\subsection{Care Respect and AI Systems}\label{sec:care}
A major contribution of care respect to the design of AI systems is as a means of introducing a broader `ethics as care' approach. While some work does explicitly appeal to the ethics of care in AI ethics debates (e.g.~\cite{asaro2019ai}), this is arguably in contrast with dominant forms of ethics in AI debates~\cite{hagendorff2020ethics}. Despite not being well integrated, the ethics of care is influential in much work approaching technology and justice from a feminist perspective (e.g.~\cite{d2020data,kitchin2020slow,taylor2020price,luka2018re,russellmaking}), as well as in the context of care robots (e.g.~\cite{vallor2011carebots,yew2020trust,johansson2013robots}). We can see how AI systems may violate care respect by failing to attend to particular and unique individual needs that care respect emphasises. Care respect focuses on the sense in which we have unique circumstances, needs, and vulnerabilities that necessitate our mutual dependency and care. From this arises the need to consider individuals as concrete individuals with specific needs and life histories, which cannot be captured by abstract feature spaces.

Second, care respect might imply different design choices to those implied by a Kantian approach with regards to paternalism and autonomy. As expressed above, Kant's anti-paternalism would suggest that out of respect for a user's rational agency, an AI system should not attempt to override their choices---even if the system predicts that those choices are not in the user's best interests. By contrast, care respect might be interpreted as requiring the system to pursue the user's needs, even if doing so appears to conflict with the user's stated desires. However, an important caveat to consider here is that advocates of care respect are not in favor of paternalism per se. Rather, they emphasise that in an inter-dependent social system where we all must care for each other at different times in our lives, we each have the ability to `make and unmake each other'~\cite{dillon1992respect}; as such, where AI is part of such a social system, the non-interference that Kant advocates is not an option.

\subsection{AI, Respect, Structural Oppression, and Domination}
As argued by Mills and others, universalist approaches to respectful AI systems which attempt to respect people by `treating them equally' are likely to compound existing inequalities of respect. This can happen when designers assume that people have equivalent experiences and expectations around the giving and receiving of respect, or fail to recognise how marginalised groups have previously suffered from disrespect and been coerced to give undue respect to dominant groups. Where this influences the distribution of features and labels in an ML system used to distribute benefits and burdens among a population, these inequalities of respect could result in distributive injustice.

At the same time, there are concerns about AI systems which have the potential to stereotype, denigrate, under-represent, and erase certain groups. In contrast to the distributive paradigm, which arguably underpins many of the definitions of `fair' machine learning models, we might also consider whether AI systems cause harms of recognition or representation. These types of harm are not exclusive: an ML model might be distributively unjust in the sense that it doesn't allocate benefits and burdens fairly, and also, separately, disrespectful in the ways that it under- or mis-represents social groups, cultures, and identities in harmful ways ~\cite{binns2018fairness}. Another term for this is `representational harm' (see~\cite{crawford2017problem}), and examples include ImageNET labelling a photo of a child as ``failure, loser, non-starter, unsuccessful person''~\cite{crawford2019ExcavatingAI}, negative stereotyping in search engine results~\cite{sweeney2013discrimination,noble2018algorithms}, and language models which replicate gender biases~\cite{bolukbasi2016man,caliskan2017semantics}.

\subsection{Systems Enacting Respect and Ascribing Agency}
A further problem with AI systems automatically inferring when to help a person is the conflation of \textit{needing} help with \textit{wanting} help. Reasons that individuals might not want help could be many and varied, including their self-identity; for instance, people may or may not identify with particular labels, such as a being disabled or having diminished autonomy, and may thus wish to exert (and demonstrate) this autonomy to themselves and others. Similarly, people may not wish to use features designated or labeled as accessibility support for such groups---e.g. not identifying as partially sighted but still wishing to use larger or heavier fonts---and may choose not to use them if these are labelled as `for the visually impaired'. Another example of how systems might ascribe agency to their users is by not forcing them through a tutorial when using a piece of software for the first time. Not doing so denies the user space to skilfully be themselves, assuming a lack of prior knowledge or ability to figure it out for themselves.

Chatbots also represent an important example of AI agents mimicking the ways that respect is enacted between humans. The performance of gender by domestic voice assistants such as Alexa and Siri clearly involves social rituals and positioning. It has often been observed that the majority of commercially available voice assistants have female voices as the default or only option~\cite{10.1145/3290607.3312915, west2019blush}, and that the positioning of these products as assistants reinforces harmful gender stereotypes. More recently, the responses of voice assistants to harassment has come under scrutiny. Against a backdrop of workplace harassment against women, trans, and non-binary people that is often inadequately dealt with by employers, and the prevalence of majority-male perpetrated physical and sexual violence in the home and the community~\cite{koss1994no}, it is important that designers acknowledge the ways that systems participate in enactments of respect and potentially perpetuate wider societal problems. By having the demeanour of ``obliging, docile and eager-to-please helpers''~\cite{west2019blush}, voice assistants reinforce the harmful aspects of social hierarchies. This way of ``presenting indirect ambiguity as a valid response to harassment''~\cite{fessler2017we} turns what may have seemed like a broadly appealing design decision---to make such devices meek and inoffensive---into an enactment of disrespect across a wide range of people beyond users of the device itself, trivialising oppressive behaviour and the marginalisation of women by a patriarchal society.

\subsection{People Respecting AI Systems}
In a similar vein, we finally consider the possibility of people paying respect to AI systems. Whether or not AI systems are moral actors, users nevertheless express respect and disrespect for these systems, and systems might be shaped to make this more or less likely, presenting both opportunities and dangers for design.

Certain kinds of respect apply naturally to systems as objects of respect. One might have a cautious respect for an immigration or similar system that makes other important decisions in the same way that one might respect a wild animal as a potentially dangerous obstacle. Attitudes like this can help convey the presence of potentially unknown risks or harms (e.g. machine learning classifiers with unknown rules or weights~\cite{seymour2020siri}). A dramatically different scenario are toys and robots designed to be the recipients of care respect, of which the Tamagotchi\footnote{\url{https://en.wikipedia.org/wiki/Tamagotchi}} or Paro~\cite{6650427} might be considered precursors, which need to be looked after for entertainment or therapeutic effects. The ethics of introducing inanimate objects designed to solicit nurturing, caring behaviours, and human emotional connections has been the subject of much ethical debate, due to such actions being seen as exploitative~\cite{wilks2010close, 6313590} and harmful, due to machines not being able to genuinely have feelings in return.

These scenarios can be understood through early work by Reeves and Nass on the paradigm of \textit{computers as social actors}, which demonstrated how people treat interactive systems as if they were people even though they know that they are not~\cite{nass1994computers, reeves1996media}. The `magic word' please-and-thank-you mode added to Amazon Alexa appears to be an extension of this, where children are asked to perform \emph{deferential} respect towards a smart speaker. Work is ongoing about the extent to which users personify these devices (e.g.~\cite{purington2017alexa}), so there remains the question of whether users are performing respect towards the physical object, its imagined personality, the organisation behind the assistant, or some combination of these. These present designers with opportunities for manipulation: inculcating emotional attachments between users and systems, and exploiting the deference that naturally arises; as well as for education: if a voice agent can detect `she' is being disrespected by a misogynistic user, `she' can answer back to challenge it.

\subsection*{Summary}
This section has taken the range of thought on respect introduced in Section~\ref{sec:background} and explored how it might relate to and inform the design of AI systems. We now develop these ideas into the concept of complex ecologies of respect, examining the different ways that AI systems shape our interactions and the potential impacts of respectful design for each.

\section{Developing Complex Ecologies of Respect}\label{sec:ecologies}
Having explored how conceptualisations of respect can be applied to different aspects of AI systems, we now develop this perspective further by broadening the scope beyond individual systems. Just as hostile architecture manifests a lack of respect for certain sections of the population by creating situations that interrupt their practices and deter their presence, datasets, models, and systems can manifest a lack of respect for individuals and groups, both in interactions where that interruption occurs and in the broader project that allows these developments uncritically. Machine vision algorithms are increasingly asked to make decisions about whether a collection of pixels are a human being. If they are unable to recognise people with dark skin tones as human~\cite{chenHPInvestigates}, they are unable to accord to them the respect and status that goes with personhood, whether this is an intelligent webcam keeping their face in frame, or a self-driving car avoiding them on a street.

\subsection{Computer-Mediated Respect: People Respecting Each Other \emph{Through} Systems}
When considering respect in the context of HCI it makes sense to consider respect as an enacted concept that arises out of action, in order to enable discussion about devices behaving respectfully towards people rather than requiring AI systems to possess the same abstract moral capacity we might believe is bestowed by personhood. By adopting the conceptualisation of respect-as-action, we must examine the relationships that exist between AI systems and people. Beyond the initial cases of devices respecting their users, and developers respecting users through devices, there are many different relationships that arise through the diverse and varied usage of AI systems in the world today (see Figure~\ref{fig:respect-configurations}). 
In the remainder of this section we will explore and discuss one of these relationships, or subject-object \textit{configurations} of respect and discuss how it might guide designers and HCI practitioners.

As well as respecting people through the way they design interactive systems, designers can also mediate the ways in which users can be respectful (or not) towards each other \textit{through} the system. The use of AI systems layered on top of digital communication channels alters the way we communicate with one another by shaping and suggesting what people say. Software such as Gmail Smart Compose\footnote{\url{https://www.blog.google/products/gmail/subject-write-emails-faster-smart-compose-gmail}}, for example, intervenes in personal communications by creating automatic replies to messages. Unpicking this example surfaces several moments where respect may or may not be shown:

\begin{itemize}
\item The system, or designer, attempts to respect the user's time, by taking the hard work out of responding in predictable ways (deference);
\item offering canned choices undermines the user's personal choice and autonomy by limiting their space to skilfully construct a response to the recipient (ascribed agency);
\item the system attempts to help the user show respect for others, by allowing them to respond quickly and indicate attentiveness (deference and social rules, although the success of this is questionable);
\item the system implicitly frames the task of replying as being a burden (obstacle respect), rather than a situated response to the actual person on the other end (care respect);
\item the system attempts to match writing styles (social rules and belongingness), although always at the risk of errors, and may go as far as following specific norms (institutional respect).
\end{itemize}

All of these connections are axes where looking at the interaction design through the lens of respect has the potential to prevent mistakes and uncover undesirable effects before they happen. Having at hand an array of different versions of respect helps to make sense of complex interactions between combinations humans, groups, technologies, infrastructures and objects. This extends the idea above that systems can perform facework towards their users, as well as towards others on their behalf. Beyond examples in messaging, we can imagine the considerable benefits and risks of future systems that mediate facework across language and cultural divides~\cite{ting1994challenge} where misunderstandings and similar failures are much more likely. When systems engage with the wide variation in values and practices between cultures, respect may offer a lens which can help motivate a more involved localisation process; instead of considering isolated practices and behaviours, conceptualisations of respect may offer insight for navigating the wider landscape of acceptable and unacceptable social interactions.

\subsection{Respect Across Boundaries}
\begin{figure}
    \centering
    \includegraphics[width=0.4\textwidth]{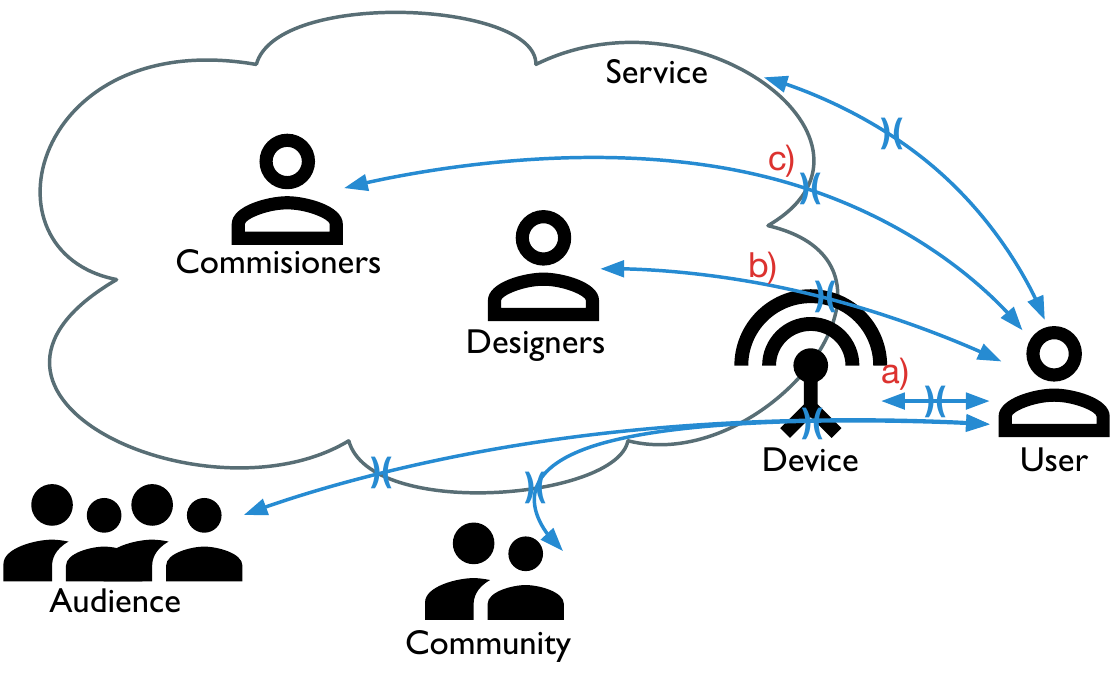}
    \caption{Configurations of respect. Each line represents a conceptualised interaction, with a particular agential cut marked. There is space for thinking about a) how the device performs respect to the user and vice versa; b) how the designer's respect or lack thereof comes through to the end user in interactions; c) how the organisations treatment of the person shows their objective respect and so on.}
    \label{fig:respect-configurations}
\end{figure}

The boundaries of computational systems are often difficult to draw, as they are rooted in a variety of materials and practices: using Alexa as an example, \citeauthor{joler2018AnatomyAI} highlight the extent of a single system~\cite{joler2018AnatomyAI}. This follows a sense from Suchman [p. 263]\cite{suchman2007HumanMachine} that a human-computer interface is not `an a priori or self-evident boundary between bodies and machines [but] a relation enacted in particular settings and one, moreover, that shifts over time', or Kitchin's critical unpacking of the nature of algorithms~\cite{kitchin2017ThinkingCritically}. The shifting nature of data-driven systems further blurs any boundaries between design and context, requiring an understanding of the ways that components are fluidly brought together and co-evolve~\cite{coulton2019MoreThanHumanCentred,giaccardi2020TechnologyMoreThanHumanDesign}. Agential realism uses the notion of \textit{agential cuts} to work with this idea that the entities under discussion are not fixed, but rather we need to draw a boundary and look at what happens across this boundary. This relates to a key concept of agential realism: that the world can be divided and compartmentalised in various ways, and the choice of division determines the phenomena observed. In placing these cuts, Shotter writes that ``we do not uncover pre-existing facts about independently existing things; we ourselves bring such `things' into existence''~\cite{shotter2014agential}. Rather than asking whether the bridge is racist, we can ask whether the bridge and the systems that created it are performing racism towards a particular group of people.

When looking at the disrespectful classifications of ImageNET, it is somewhat difficult to pinpoint exactly where disrespect occurs in the constellations of actors and actions, the repurposing of various data, and this is worthy of further analysis (Section \ref{sec:repurposing}). When working with a connected system, there are many places where lines could be drawn: one could separate the user from the device and all the services behind it in order to look at how a smart speaker performs respectful interaction; another cut would look at how two users interact through the system and develop notions of computationally mediated respect and so on. An embedding of a voice assistant in everyday life ~\cite{porcheron2018VoiceInterfaces} suggests multiple cuts that can give rise to different pictures of respectful interaction (Figure \ref{fig:respect-configurations}).

This tends to fit with the glosses made when talking about AI systems: asking about whether Alexa is sexist~\cite{womenReclaimingAI} is asking a question that includes the moral agency of the designers and programmers, the data and training sets as well as the way that it is advertised and situated as a device to have within a home. 
At the same time, increasing intelligence and interactivity in such systems supports a view that they have agency: they respond, shape, intervene, choose, and so on. All this can be granted without necessarily having to discuss the prospect of giving them rights or personhood (often a distraction from the more pressing needs of those whose welfare they harm~\cite{birhane2020robot}), using respect as a unifying concept to look at the actions and understandings made.

\section{Developing Respectful AI Practices}
As separate but related concepts, the conceptualisations of respect discussed above are clearly not always concordant. As we have seen, there are often major theoretical differences between approaches. For this reason, arriving at a purely philosophical definition of respect for AI systems is likely to be contentious---for example, consider the vast differences in underlying assumptions between the Kantian deontological approach to respect for persons and the flat ontological structure of agential realism.

But there are many areas where we believe agreement is possible; or at least, coherence around sets of ethical concerns and perspectives which respect in its various guises can help designers to navigate. In this section, we discuss how respect can be applied to different stages of the design process and how respect extends the ethical obligations of designers to consider how their products allow users to treat each other, before finally discussing the similarities and differences between respect and fairness in more depth.

\subsection{Respect as a Lens for the Design of AI Systems}
Respect is a lens that can---and we argue should---be brought to bear on all aspects of a system's life-cycle, from the ways that data is collected and the organisational philosophies behind the choice and orientation of projects through to the micro-interactions that shape the relations between individual end-users and their devices. At each stage, there are possibilities for disrespect, marginalisation and a wide selection of harms to be accidentally inflicted; failures of respect early in the process can be carried through and magnified as components are assembled into systems. 
When considering existing design approaches in HCI, we might ask how frameworks such as user-centred (UCD) and value-sensitive (VSD) design relate to the respectful design of AI systems. For instance, in centring users and prioritising their needs, UCD could be seen as embodying a version of respect for persons articulated by the categorical imperative by treating users as ends in themselves. But at the same time, common practices within UCD, like the use of personas to direct the design process towards the unique needs of a system’s users, necessarily abstract away from individuals by representing a large number of people as a small number of personas with definable characteristics. This presents the dual risks of using stereotypes rather than archetypes, and introducing diversity of gender, ethnicity, and age to personas where this is ``unimportant to the design [process]''~\cite{cooper2014face} without knowledge of current and historic injustices, furthering the idea of a colourless, genderless design subject.

With value-sensitive design, we see respect as a powerful tool to help narrow down the potentially infinite selection of values that could be designed for in any given system. There are a number of values that underpin the different kinds of respect described above, such as equity, autonomy, identity, and care, and exploring which kinds of respect are most relevant to the situation at hand therefore leads to a much smaller and well defined set of candidate values. In this sense, respect is not (just) another collection of values to be considered within VSD, but a way to structure and relate to other values. When conflicts occur between different kinds of respect in a system, it elucidates the \textit{why} behind difficult design considerations and helps to show what is at stake in a decision. One example of how this could manifest is in devices designed to support ageing in place that might impinge on users' autonomy and privacy~\cite{demiris2008senior}; consider a smart thermostat that prevents the user from setting the temperature dangerously low (perhaps in an effort to save money). The conflict between Kantian non-paternalism, care respect, and the ascription of agency allows for designers to clearly weigh up decisions about the development of a product with regard to the ethical issues under discussion.

Other approaches such as inclusive and universal design are directly targeted at addressing the forms of disrespect uncovered in earlier sections, and are thus closely aligned with many of the types of respect we detail above. As well as further motivating the need for more inclusive design approaches, respect helps by clearly distinguishing the different aspects of social and cultural inclusion that drive these inclusive approaches, providing a clear way forward towards their realisation. One of the closest approaches to the current work is on designing for human rights in AI~\cite{aizenberg2020DesigningHumanRights}, which looks at ways to bridge the gaps between abstract value language and clear design requirements and processes.

Returning to previous uses of respect in HCI, it is now clear how our development of respect relates to and extends prior work. Formulations of respectful design that take into consideration the unique values, motives, and perspectives of users~\cite{marti2014in, rajapakse2019RespectfulDesign} clearly relate to care respect as described in Section~\ref{sec:care}. Approaches prioritising privacy, security, and anti-abuse~\cite{kissner2019respect} are rooted in the same fundamental principles as the categorical imperative, describing the treatment that all users deserve from AI systems. By mapping out the landscape of respect as it is most relevant to HCI we hope to place these existing efforts on a firm philosophical footing and provide the tools required for researchers and practitioners to continue to integrate respect into their creations.

\subsection{Respect During the Systems Development Lifecycle}
We now consider examples of how respect represents a crucial lens when gathering initial requirements, data collection through to deployment and beyond. We hope that respect can offer a pragmatic viewpoint here---where fairness is grounded in mathematical properties and ethics in philosophical systems, respect can take a more personal, enacted and pragmatic view, tending towards engagement and understanding of actual humans rather than abstracted ideas. Rather than checkboxes, respect can supply a productive set of questions that underpin a practice that can be enacted all the way through: what is a respectful way to collect or process data? What is a respectful way of interacting with users and stakeholders?

\subsubsection{Creating New AI Systems}
When determining the initial requirements and outline of a system being commissioned, the paradigms chosen will shape the way that the system might perform (dis)respect to those it ultimately affects. Using classification techniques for a college admission or criminal justice AI system, for example, precludes the individual treatment of the people who will be judged by the system and requires the use of quantifiable features as proxies for intangible (and often morally questionable) qualities like intelligence and criminality. Regardless of whether these systems are subsequently engineered to be fair, the high level decision to make life changing decisions about people based on a limited number of traits fails to show care respect---``recognizing our power to make and unmake each other as persons''~\cite{dillon1992respect}) and are likely to perpetuate forms of structural oppression by considering the experiences of people with similar features to be the same (see Section~\ref{sec:individual-respect} below). Similarly, the act of representing knowledge or creating systems without the involvement of the community is fundamentally disrespectful, as what Dourish calls the `colonial impulse' of ubiquitous computing~\cite{dourish2012UbicompColonial} undermines the ways that people construct their own identities~\cite{bidwell2016MovingCentre} and knowledge~\cite{winschiers-theophilus2010DeterminingRequirements}. This can be seen also in projects such as incorporating favelas into the overarching data object of Google Maps -- it fails to respect the social practices and desires of many of the people living there, or their desires for obscurity~\cite{luque-ayala2019DigitalTerritories}.  This is a space where respect---for people, cultures, practices---can offer a structure for a more considered design and commissioning of systems~\cite{kotut2020ClashTimes}. 

\subsubsection{Collecting and Re-purposing Data}
\label{sec:repurposing}
ImageNET has been held up as an example of datasets gone wrong~\cite{crawford2019ExcavatingAI,pasquinelli2020NooscopeManifested}, as images are labelled with sexual slurs and dehumanising language. A question here, then is how a greater emphasis on respect could have helped to avoid this kind of problem. The original WordNet categories are descriptive of language terms, and so not particularly problematic beyond the prioritisation of a particular view on language and its use. However, when moving from descriptions of language to descriptions of \emph{people}, it is clear that a synonym relation between \emph{poor} and \emph{pathetic} or \emph{piteous} is disrespectful -- it treats people as caricatures of a certain categorisation, applying an evaluative disrespect through equating lack of money with lack of ability alongside a lack of concern for individual circumstance.
The use of social media images to construct models to classify e.g. sexual orientation~\cite{wang2018DeepNeural} or political persuasion~\cite{kosinski2021FacialRecognition} raises many concerns~\cite{pasquale2018WhenMachine}; simplifying the questions to ones rooted in respect for persons and recognition respect -- ``would the person in this image want to be classified this way'', ``how does this classification affect the representation of these groups'' -- offers a pragmatic approach to avoiding harm, representational or otherwise.
The possibility for respectful treatment extends to the processing of the data as well -- for ImageNET, workers on Amazon's Mechanical Turk were employed to carry out the classification. This kind of atomised practice, that separates workers from each other, and separates the data from its context similarly separates the activity from its implications. A more respectful processing of data would include the people that the data was about in the process, as those who know themselves best, giving the possibility of shaping or objecting to the process.
As datasets are processed, broken up, combined and brought together into complex ecologies (Section~\ref{sec:ecologies}) there are clear possibilities for systems where individual decisions and actions are reasonable, but the overall effect fails a check for respectfulness. This viewpoint helps to capture situations where data uses are at odds with the circumstances of its collection, or feelings like the `creepiness'~\cite{pasquale2018WhenMachine} of inferring protected characteristics from images.

\subsubsection{Respectful human-algorithm interaction design}
When it comes to the interaction between users and AI systems, respect provides a simple framework for thinking about consequences. Alexa's early responses to sexist language were highlighted as problematic~\cite{fessler2017we}---an embodied micro-interaction that highlighted systemic disrespect. This is not entirely surprising---a system developed in a culture that does not act respectfully is unlikely to be respectful; an engineering team where women are well represented is unlikely to prioritise the creation of `catch-me-if-you-can flirtation' and other male fantasies~\cite{walker2020AlexaAre, west2019blush}. Beyond questions of why a female voice is appropriate in the first place, this is an example of how those who are not at the table can be casually disrespected by only representing the convenient parts of their gender identity. An example of a less embodied disrespectful micro-interaction would be algorithmic content moderation, where certain topics or types of image are technically disallowed. Understanding the relations between content moderation and users highlights where the algorithmic viewpoint leaks out, and narrow attempts to shape user behaviour lead to unintended and discriminatory outcomes as components come together into e.g. `sexist assemblages'~\cite{gerrard2020ContentModeration}. 

Respect is particularly useful for interrogating these micro-interactions, as it can better capture the kinds of harm engendered -- the elisions, erasures and smoothings that reduce humanness; the universal, default voices and viewpoints that demand conformity are hard to address with fairness or transparency, but brought into sharp relief when asking whether interacting with this person, in this way, with this voice is a respectful thing to do.

\section{From Fairness to individual respect}
While issues of societal bias, discrimination, and fairness in sociotechnical systems have long been studied in a range of disciplines~\cite{friedman1996bias,gandy1993panoptic,hutchinson201950}, the last decade has seen a significant growth of interest in these topics amongst computer scientists, often borrowing from or in collaboration with other disciplines such as law, philosophy and social sciences~\cite{pedreshi2008discrimination,dwork2012fairness,barocas2016big,binns2018fairness}. Alongside transparency and accountability, fairness has emerged as a key design goal for machine learning systems~\cite{chouldechova2020snapshot}. Like respect, fairness is a concept which admits multiple interpretations across a variety of disciplines. Some of these have been implicitly or explicitly appealed to as the philosophical grounding for various statistical fairness metrics proposed within recent fair machine learning literature.

A key difference between fairness and respect in these contexts, as discussed above, is that the former generally concerns individuals in so far as they are members of groups, whose outcomes are compared, whereas many conceptualisations of respect focus on the treatment of individuals. Even concepts like individual fairness still essentially equate an individual with their position in the feature space of a machine learning model, which is based on generalisations from other people in the training data~\cite{binns2020apparent}. In this sense, individuals are not treated as individuals, but on the basis of generalisation; and arguably, as means to ends (in this case, the end of achieving a `fair' model). As such, they may violate, or at least fail to satisfy, the Kantian notion of respect as the genuine treatment of people as ends in themselves rather than solely as means, and the care respect notion that we each have unique needs and circumstances. As a contrast to the computational notion of individual fairness, then, AI system designers might instead try to show `individual \emph{respect}', and we now explore what this might mean.

\subsection{Developing Individual Respect}\label{sec:individual-respect}
It is now common for many kinds of AI systems to classify and differentiate people based on properties such as their demographics, preferences, activities, and so on. The fairness literature has identified issues around fairness and minority and marginalised groups: if we are to test that an algorithm affords the fair treatment of people from different groups, is a group ever too small to be considered?

This perspective relates to the concept of \emph{intersectionality}. Originating in work by Black, Latina, and Native women in the 18th century~\cite{collins2020intersectionality}, intersectionality highlights the ways that existing anti-discrimination measures tend to treat race and gender as discrete categories~\cite{crenshaw1989demarginalizing}. Recognising that people are more than a collection of disparate categories---that experiences of race, age, gender, ability, and so on interact with each other---some fair machine learning research has proposed intersectional fairness measures rather than `single-axis' measures which consider protected characteristics individually~\cite{kearns2018PreventingFairness, buolamwini2018gender}. Noting the large number of possible combinations of characteristics, and the possibility that other features could be proxies for them, one might be tempted to take this notion of fairness to its logical conclusion and treat every possible configuration of features as a sub-group in its own right and test for disadvantage. Where such sub-groups contain only one individual, intersectional fairness begins to look like a kind of individual fairness. 

However, even if statistically and computationally tractable, this understanding of intersectionality is limited. As Hoffman argues, intersectionality is not ``a matter of randomly combining infinite variables to see what `disadvantages' fall out; rather, it is about mapping the production and contingency of social categories''~\cite{hoffmann2019fairness}. In practice, both intersectional fairness and individual fairness struggle to account for the inherent limits of algorithmic classifiers---two people with the same characteristics will always receive the same outcome, and thus have their situations considered equivalent, even though there must necessarily be other aspects of their lives that are not represented within the classification system.

There are multiple ways this could be seen as disrespectful; firstly, because it suggests in some sense that people are defined by, or are similar because of, such attributes, rather than reflecting on the ways people are unique and individual. Embracing individuality may mean systems should be sensitive to the individual experiences people have, due to there being no singular account of what it means to be a member of any category, such as being black, a woman, an older adult, or to experience hearing loss or a loss of mobility. Secondly, such a categorisation overlooks the routes that people take towards the position they occupy in feature space, such as the personal histories, oppressive structures, traditions, or environments that constrain or shape them.   Thus, a supportive and respectful mediating system has the responsibility of considering these factors, structures, and histories when determining the appropriate ways they should respond.  However, there are inherent limits to AI systems in this regard, because they are incapable of considering factors that lie outside the input features, or beyond the `feature horizon'; we might consider this an instance of, or analogous to, the `frame problem' that has plagued AI research since its inception~\cite{shanahan2006frame}.\footnote{Having a `human in the loop' might ameliorate this concern to some extent, but would only assure respect for every individual affected by the AI system if the human in the loop genuinely has the time and care to consider possible differences lying beyond the model's feature horizon. Such a degree of human attention and care would typically run counter to the common rationales of cost-saving and efficiency which underly the deployment of AI systems in the first place~\cite{binns2019human}.}

\section{Conclusion}
We have attempted to bring the concept of respect to the point where it can be used meaningfully in discussion about the design and analysis of AI systems. This is complicated by the fact that respect is a varied concept, with multiple formulations and areas of concern. However, the broadness of respect as a lens is also its strength: it brings together a collection of important ideas, in a way that lines up with communicable, commonsense reasoning, that can be applied at any stage of a process. 
At the start of system design, asking questions around respect deals with fundamental views on the way that people can or should be represented, classified or engaged with. This allows for some radical rethinking of what AI systems are and do, and how they relate to the people which they model and decide for. In particular, respectful design can help with challenging the default positivist views around representing the world through data, models and generalisation.
As a system comes together, respect can help to look across ideas such as fairness and transparency, and consider issues of representation, equality, treatment and agency. It connects between fairness, that plays out at population level, and personal experience, while offering a pathway to bringing in more ethical behaviour, and a framework to both inform and critique practices such as USD and VSD.

\begin{acks}
This work was part-funded by grant EP/T026723/1 from the Engineering and Physical Sciences Research Council.
\end{acks}

\bibliographystyle{ACM-Reference-Format}
\bibliography{main}

\end{document}